# Theoretical and experimental search for ZnSb based thermoelectric materials

**K Niedziolka, R Pothin, F Rouessac, R M Ayral and P Jund**
Institut Charles Gerhardt, Université Montpellier 2, Pl. E. Bataillon CC1506 34090 Montpellier, France

E-mail: pjund@univ-montp2.fr

**Abstract.** We report a combined theoretical and experimental search for thermoelectric materials based on semiconducting zinc antimony. Influence of three new doping elements (sodium, potassium and boron) on the electronic properties is investigated as well as the carrier concentration and temperature dependence of the thermoelectric coefficients obtained through density-functional calculations and Boltzmann transport theory. Distortion of the electron arrangement caused by the doping elements is displayed as a deformation charge density around the atoms. Based on the band structures, the density of states and the transport properties, we find that the presence of Na and K in the ZnSb matrix leads to a slightly improved p-type conductivity while the B substitution leads to a n-type doping. Due to the stronger need of obtaining a n-type ZnSb based material, the $B_{0.01}Zn_{0.99}Sb$ structure has been transferred to the laboratory in order to be synthesized by direct melting. The sample is investigated by X-ray diffraction and Scanning Electron Microscopy.

1. **Introduction**

The thermoelectric effect is nowadays a widely discussed subject. Thermoelectricity can make a contribution that would not only accelerate the reduction of emissions of greenhouse gases including in industrial systems and transport, but also allow greater autonomy of portable electrical appliances. Unfortunately, the application of thermoelectric materials has been limited mainly due to their low efficiency. This property can be quantified by the dimensionless figure of merit, $zT = \sigma S^2 T/\kappa$, where $\sigma$ is the electrical conductivity, $S$ is the Seebeck coefficient, $T$ is the temperature and $\kappa$ is the thermal conductivity. In the middle of the 20[th] century the thermoelectric effect was the subject of work of several scientists[1, 2] but they weren't able to synthesize materials with a high figure of merit. The best discovered materials for thermoelectric applications have been alloys based on bismuth and tellurium[3], with a figure of merit around 1. The second life of thermoelectric materials begun in the 1990s, when a high performance has been theoretically predicted (for example for nanostructured materials[4]). Simultaneously experiments on complex bulk materials have been performed revealing some compounds (e.g. clathrates [5], tellurides – LAST and TAGS[6], skutterudites [7] and Zintl phases[8]) with a high figure of merit. Among the latest compounds stoichiometric zinc antimony distinguishes itself due to the high abundance and the low cost and toxicity. What is important as well, ZnSb is stable in the 300-600K temperature range for which only few efficient thermoelectric materials have been found. Therefore ZnSb can be considered as one of the best candidate materials in this temperature range, but to achieve performances greater than 10% required for industrial use one should further improve its thermoelectric properties, and one way of doing it is by doping.

In order to manufacture a thermoelectric device both p- and n-type modules are necessary. Furthermore both modules should be based on the same compound if one wants to avoid the problems associated with the thermal expansion of two different materials. So far ZnSb with stable properties has been found only of p-type. The ZnSb 1:1 phase has been reported to have a $zT$ around 0.2[2] and 0.6[9] at 273K and 460K respectively. Recently, an attempt for increasing the figure of merit by decreasing its thermal conductivity by ball milling and microstructuration has been made[10]. Nevertheless, no improvements of the thermoelectric properties have been achieved. Furthermore, despite numerous attempts no stable n-type doped ZnSb has been found yet. Several p-type doped ZnSb structures have been reported: Al[2], Cu[2, 11], Sn[11], Cr[12], Mn[12], Ag[10, 13]. Also a couple of structures have been reported as n-type, namely the ones doped with Te, In and Ga[14] . However all of them show a high instability and convert to p-type over the whole temperature range. Moreover unstable n-type ZnSb has only been synthesized as single crystals. Since n-type ZnSb is essential to produce efficient thermoelectric modules we continue the search. Firstly we have simulated by *ab-initio* calculations, the electronic, thermal and thermoelectric properties of pure ZnSb and of some doped compounds to determine the best material with a theoretically predicted thermoelectric figure of merit. As mentioned above this factor is proportional to the electrical conductivity, and inversely proportional to the thermal conductivity. Doping is necessary to optimize the electronic conductivity and the thermoelectric power by making junctions, while nanostructuring is essential to reduce the thermal conductivity without affecting the electrical conductivity. In the present paper we show the results of attempts to dope ZnSb with Sodium, Potassium and Boron, elements that have not been considered before. Furthermore since our aim is to produce an n-type ZnSb based thermoelectric material, we need to dope the ZnSb matrix with elements providing electrons to the structure. The natural choices are alkali and alkali-earth elements as well as elements at the right of Zn in the periodic table if we want to find dopants on the Zinc site. In addition it is worth mentioning that Sodium and Potassium are creating with Antimony the intermetallic compounds NaSb[15] and KSb[16], while boron is creating the covalent compound BSb[17].

This paper is organized as follows. In section 2 we describe the calculation procedures and the details of the synthesis and characterization methods. Theoretical and experimental results and a discussion concerning structural, electronic and transport properties are presented in section 3. The paper is concluded in section 4.

## 2. Procedures

### 2.1. Computational details

Theoretical simulations were performed by means of the first-principle projector augmented waves (PAW)[18] method, as implemented in the highly efficient Vienna Ab initio Simulation Package (VASP)[19]. Exchange-correlation effects were treated within the General Gradient Approximation (GGA) with the Perdew-Burke-Ernzerhof functional (PBE)[20]. The plane-wave energy cut-off was fixed to 500 eV and kept constant through all the calculations. Brillouin zone integrations were done within the GGA method on a Monkhorst-Pack 3x3x3 k-points grid for the relaxation and a 5x5x5 grid for the densities of state and the charge transfer calculations. In all of the calculations the convergence limit was set to $1 \times 10^{-4}$ eV/Å for the forces and to $1 \times 10^{-5}$ eV/Å for the energy. For pure 2x2x2 super cells we have performed the structural optimization by minimizing the total energy with respect to the cell parameters and atomic positions, and afterwards we have maintained the cell parameters and allowed the atomic positions to relax for the doped structures $AZn_{63}Sb_{64}$ (A=Na, K, B).

Charge transfers were calculated using the Bader Charge Analysis[21, 22]. Following the recipe from ref[23]to ensure a high accuracy of the charge calculations, the mesh for the augmentation charges was tested starting from the mesh size of the relaxation calculation, and increasing it stepwise by 50% up to 350%. A grid size of 280x336x360 (200%) was enough to secure the convergence of the charge transfer between the atoms.

The transport properties (Seebeck coefficient, Power Factor and figure of merit) were calculated using the BoltzTraP[24] program, with the Boltzmann Transport Equation (BTE) and the constant relaxation time approximation.

*2.2. Synthesis*

Pure and Boron doped samples of ZnSb (0.5at.% of boron) were synthesized by the direct melting of the constituent elements in an evacuated quartz tube. The starting materials were antimony (99.999% pure Alfa Aesar), zinc (99.9% Alfa Aesar) and amorphous boron (99.99 % Alfa Aesar). The preparation of the starting mixtures was carried out in a glove box in order to avoid oxidation. In order to get a good homogeneity of the liquid phase with the sample containing 0.5at% of boron, three consecutive melting processes were realized at 973 K, for 3h30min each and the ampoule was turned upside-down every time. Two melting processes were followed by air-quenching and the last one by water-quenching. After that, the ingots were annealed in the ampoule at 523 K for two weeks.

*2.3. Characterization*

The resultant materials were investigated by X-ray diffraction (XRD) using the Cu $K_\alpha$ radiation in the range of 2θ from 20 to 60 degrees. Before Scanning Electron Microscopy (SEM) analysis, the ingot was cut into slices. Furthermore, they were placed in Wood alloys (to prevent carbon pollution for the SEM analysis) to be polished with water using SiC abrasive paper (roughness 800, 1200, 2400, 4000) and then diamond solutions (3 μm, 1μm and ¼ μm). Some of these analyses were performed on freshly-cut surfaces (without polishing) in order to keep all available information. The observation of the morphology of the materials coupled with the quantitative chemical analysis was performed on a Scanning Electron Microscope FEI Quanta 200, resolution 3 nm, vacuum, 30kV for the secondary electrons and 4 nm for the backscattered electrons, coupled with a microprobe EDS (Energy Dispersive X-ray Spectroscopy) Oxford Instrument XMax with a detector of 50 mm².

3. **Results and discussion**

*3.1. Structural data and electronic charge distribution*

ZnSb is a II-V orthorhombic semiconductor that crystallizes within the *Pbca* space group. Calculated and experimentally obtained lattice parameters are reported in table 1. Theoretical values of the cell parameters are about 1.5% overestimated in comparison to the experimental values, which is reasonable for GGA-DFT based calculations.

Table 1. Super cell lattice parameters determined theoretically and experimentally[25].

| ZnSb super cell (2x2x2) | a | b | c |
|---|---|---|---|
| Exp | 12.4032 | 15.4832 | 16.1990 |
| GGA | 12.5616 | 15.6492 | 16.4586 |

The unit cell of this compound consists of 8 Zn and 8 Sb atoms, that can be grouped into 4-membered rhomboid rings as has been already described[26] and as it is presented further in the paper (figure 1). The properties of ZnSb have been carefully investigated in the past[27, 28] and more recent studies[13, 29]. This slightly anisotropic p-type semiconductor has an experimental band gap of about 0.53 eV[27].

The classical way to extract bonding properties is to analyze the electron density contribution. For this purpose we have determined the maps of deformation charge densities. In all these maps the reference state is a superposition of the free non-interacting atoms – its calculated charge density has been subtracted from the crystal charge density. Figure 1 displays the situation for pure and doped ($AZn_{63}Sb_{64}$) structures of ZnSb. Charge densities can be seen in the planes defined by 4 atoms that belong to the rhomboid rings. For pure ZnSb these rings are made of 2 zinc and 2 antimony atoms. For doped structures one zinc atom is replaced by a doping element (A=Na, K, B).

Considering the undoped structure, the density maxima appear on the shorter interatomic lines between Zn and Sb, slightly shifted from the center of a bond into the direction of the more electronegative Sb atom (electronegativities: 1.65 (Zn), 1.96 (Sb)). The Bader Charges reported in table 2 are giving us more details about the bonding. Small calculated charge transfers between Zn and Sb demonstrate the mostly covalent character of the Zn-Sb bonds in pure ZnSb. After doping, the substitution of Zn by sodium or potassium has distorted the symmetry of the 4 membered rhomboid rings, elongating the distance between the doping element and the rest of elements of the ring. Sodium and potassium are elements from the first group of the periodic table with low electronegativities (0.93 (Na), 0.82 (K)). Their tendency to donate an electron in order to achieve the electronic configuration of the closest noble gas can be clearly seen in figure 1 (b) and 1 (c). The electron of Na/K migrates to the most electronegative element in the system: Sb (see table 2 where the average atomic charges are displayed). Indeed when analyzing the individual Bader charges on the Antimony atoms, one notes a clear charge transfer towards the nearest Sb neighbors around the Na or K atom, since they are more negatively charged than the other Sb atoms. It is worth mentioning that due to the presence of the dopant the charge transfer between Zn and Sb is less important than in the pure compound which explains the slightly less negative charge on the Sb atoms and the slightly less positive charge on the Zn atoms in table 2. The situation looks different for the structure doped with boron. In this case the structure of the 4 membered rings is also distorted but in a reversed manner. The interatomic distances become shorter (figure 1 (d)). Also the accumulation of charge between B and Sb is obvious, creating a covalent bond between those two atoms, which is not surprising since the BSb[30] semiconducting compound has been reported with a high cohesive energy[17]. The charge transfer in this case is also reversed – it goes from antimony to the more electronegative boron (2.04). It is worth mentioning that in all of the compounds the charge on zinc is sensibly the same.

The above results are also coherent with simple valence electrons counting. The substitution of Zn by Na or K results in p-type doping through the removal from the structure of one valence electron, while the substitution by B is adding one valence electron giving rise to n-type doping.

Table 2. Integrated average atomic charges according to Bader[21, 22].

| Compound | Charge transfer (charge on atom) | | |
|---|---|---|---|
| | Zn | Sb | A |
| $Zn_{64}Sb_{64}$ | 0.2618 | -0.2618 | - |
| $NaZn_{63}Sb_{64}$ | 0.2516 | -0.2597 | 0.77 |
| $KZn_{63}Sb_{64}$ | 0.2511 | -0.2578 | 0.68 |
| $BZn_{63}Sb_{64}$ | 0.2512 | -0.2409 | -0.41 |

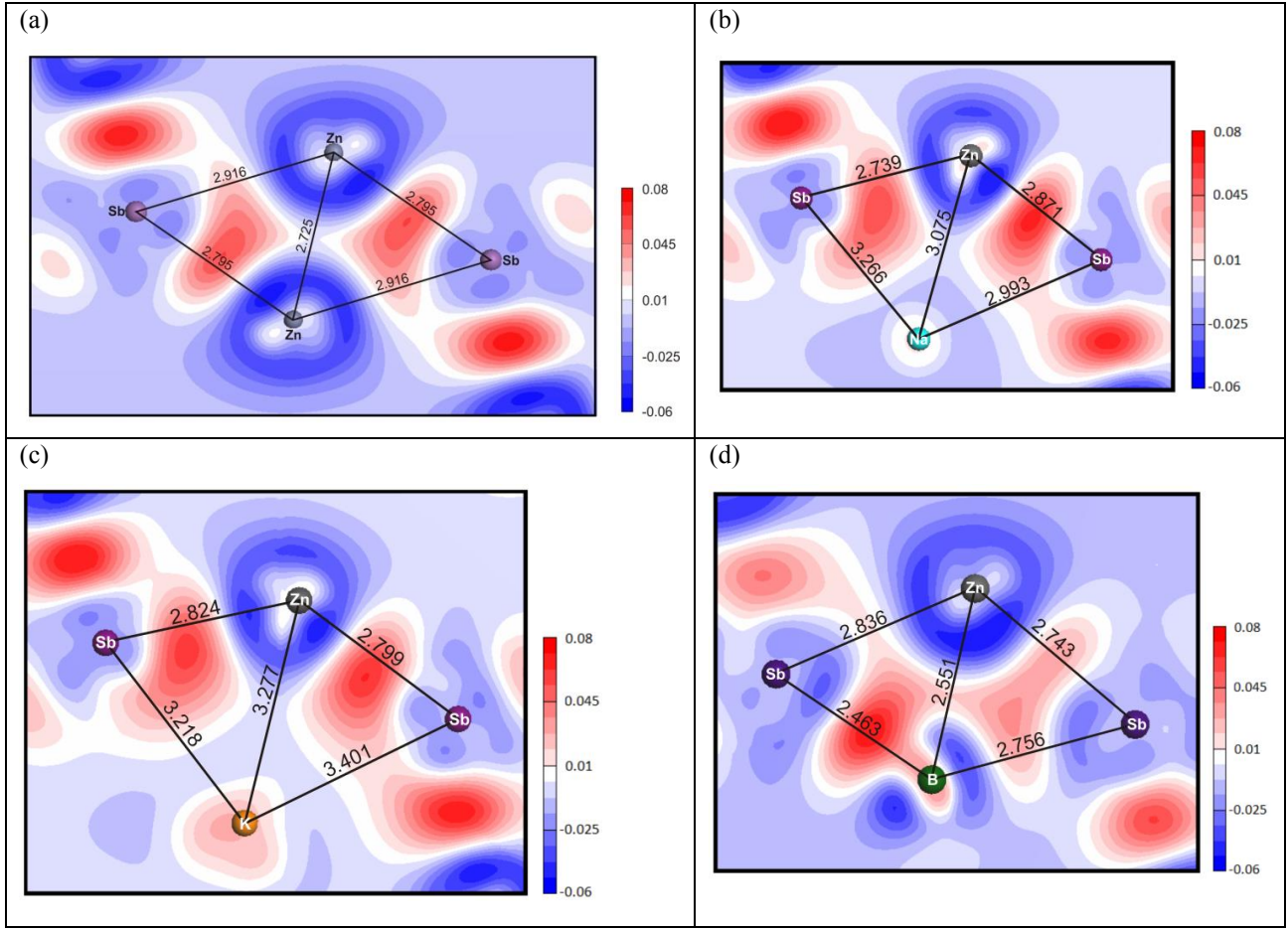

Figure 1. Deformation charge density distribution in electrons per Å$^3$ in the planes of the 4 membered rings for (a) pure zinc antimony (b) ZnSb doped with sodium NaZn$_{63}$Sb$_{64}$ (c) ZnSb doped with potassium KZn$_{63}$Sb$_{64}$ and (d) ZnSb doped with boron BZn$_{63}$Sb$_{64}$. The color map indicates the isocharge density lines.

## *3.2. Electronic properties*

The determination of the thermoelectric properties, and in particular of the *zT*, requires a very accurate electronic structure. In figure 2(a-d) the calculated density of states and band structures along high symmetry lines at 0K are shown. For all of the considered structures the absolute maximum of the valence band occurs along the *Γ-X* direction and the absolute minimum of the conduction band occurs along the *Γ-Z* direction. A similar shape of the bands and especially of the conduction and valence bands is observed. The most noticeable change is the deconvolution of degenerate bands in the doped structures. The substitutions flatten these bands somewhat and as a consequence slightly increased effective masses are expected. Larger effective masses lead to an increase of the thermopower. No additional impurity band in the vicinity of the energy gap is observed. From the DOS it is also noticeable that the edge of the conduction band in B doped ZnSb – $BZn_{63}Sb_{64}$ is composed primarily of the electronic states of boron, while the electronic states of Na in $NaZn_{63}Sb_{64}$ and K in $KZn_{63}Sb_{64}$ are dispersed more evenly. The calculated energy gap for the pure ZnSb super cell (128 atoms) is equal to 0.03 eV and is even smaller than the previously reported[26] value for the single cell (16 atoms): 0.05eV. As expected from DFT-GGA calculations - those values are very much underestimated in comparison to the experimental one- 0.53eV[27]. The evaluated energy gaps for all four structures are reported in table 3 along with the positions of the Fermi levels. The nature of the doping element does not significantly change the value of the energy gap.

Considering the structures doped with sodium and potassium, where the energy gap is above the Fermi level a notable amount of the valence band remains unoccupied. If existing, this material would be a good metallic conductor. For the boron doped structure the Fermi level lies above the energy gap. As a consequence p-type doping is predicted when inserting K and Na in the ZnSb matrix and n-type doping when inserting B. These results are also consistent with the electron counting method mentioned above.

Table 3. Theoretically predicted energy gap (eV), position of the Fermi level with respect to the top of the valence band (eV) and the effective mass ($m_0$).

|  | ZnSb | $BZn_{63}Sb_{64}$ | $NaZn_{63}Sb_{64}$ | $KZn_{63}Sb_{64}$ |
|---|---|---|---|---|
| Energy gap (eV) | 0.03 | 0.013 | 0.05 | 0.069 |
| Position of Fermi level(eV) | 0.02 | 0.14 | -0.21 | -0.14 |
| Effective mass ($m_0$) | -0.18/0.1 | 0.13 | -0.2 | -0.27 |

In figure 2 (the calculated band structures and densities of states), one can see that all of the doped structures have a metallic character, since their Fermi level lays outside of the energy gap. Nevertheless since the GGA calculations gravelly underestimate the energy gap the predicted metallic character is not certain. Analysis of the four band structures shows that doping elements in the present concentration are not changing their shape significantly. Therefore it seems reasonable to use a scissor operator for a given band structure to try to determine if the doped structures are really metallic. In order to do so we have also performed the calculations on a pure ZnSb single cell using the HSE[31]hybrid functional which leads to a theoretical band gap of 0.56 eV (details will be given in an other article). Afterwards the bands above and below the energy gap have been shifted accordingly – up and down by half of the value of the difference between the energy gap calculated within the GGA for a given doped structure and the calculated gap within HSE for undoped ZnSb. For all of the structures after opening the gap, the Fermi level lays inside the energy band gap, proving the possibility that at the considered concentration of doping elements the compounds remain semiconducting which is favorable for thermoelectric applications.

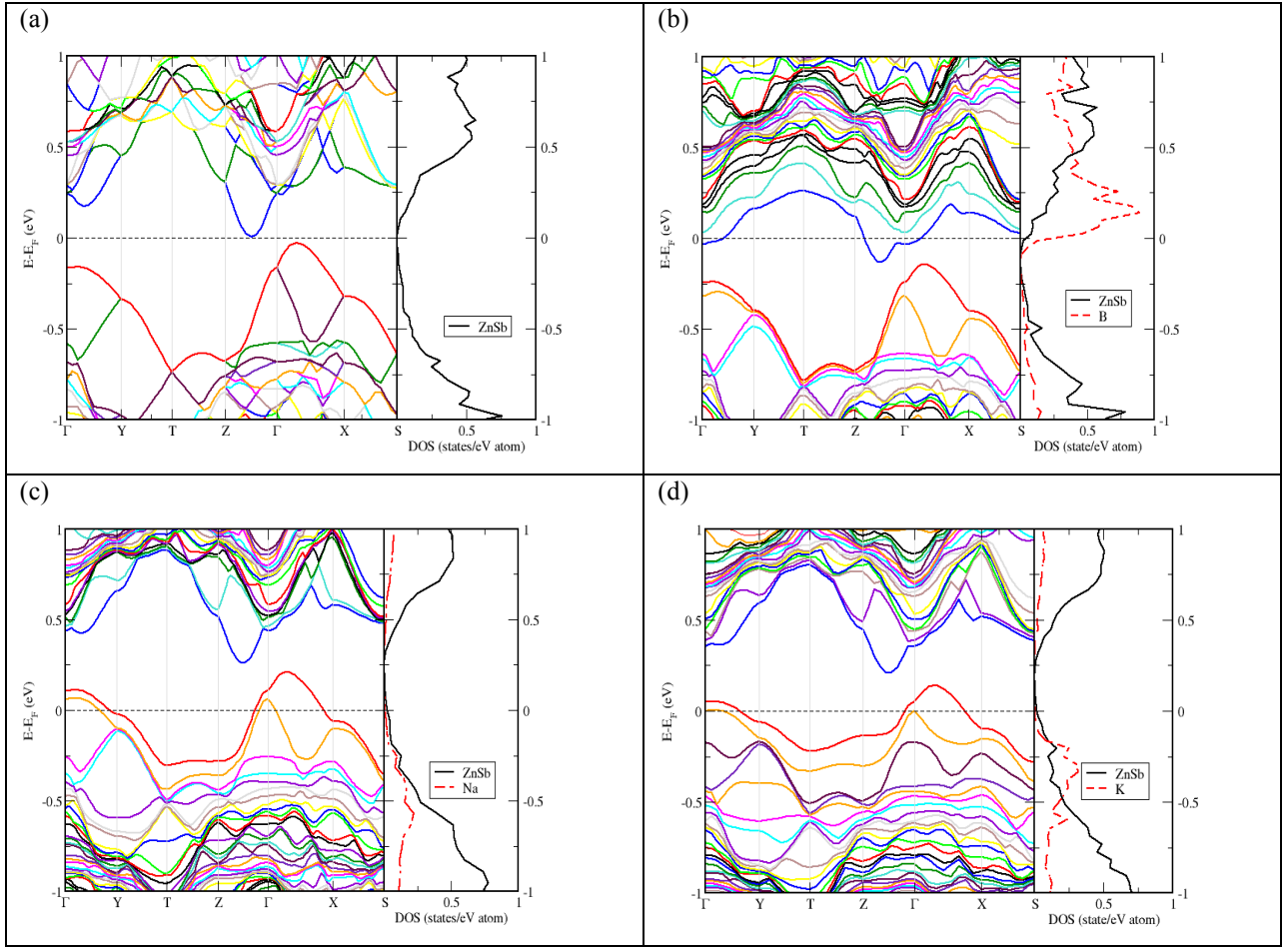

Figure 2. Band structure along high symmetry lines of: a) pure ZnSb super cell (128 atoms); b) Boron doped super cell BZn$_{63}$Sb$_{64}$; c) Sodium doped super cell NaZn$_{63}$Sb$_{64}$; d) Potassium doped super cell KZn$_{63}$Sb$_{64}$. The band energies are displayed with respect to the Fermi level.

The valence bands are nearly parabolic in the region of the valence band maxima for all the structures. Therefore the effective carrier masses have been determined by fitting the valence band at the maximum with a free electrons type equation:

$$E = \frac{\hbar^2 k^2}{2m^*} \quad (1)$$

where $\hbar$ is the reduced Planck's constant, k the wavevector and m* the effective mass.

The resulting values are shown in table 3. The experimental value of the effective mass of pure ZnSb has been determined by Komiya et al.[27] at $0.175m_0$ along the a axis (without determining the sign but assuming that this value pertains to holes). The theoretical effective mass of $-0.18m_0$ calculated for holes in pure zinc antimony (in the vicinity of the top of the valence band) agrees very well with this experimental result. The effective mass of the electrons has also been calculated (in the vicinity of the bottom of the conduction band) - the results are included in table 3 as well. Slightly higher values of -0.2 and $-0.27m_0$ have been obtained for p-type structures doped with sodium and potassium, hence enhanced thermoelectric properties can be expected.

### 3.3. Transport properties

We have calculated the transport properties by means of the Boltzmann transport equation approach. Furthermore they were calculated based on the rigid band model and constant relaxation time approximation. The validity of these approaches has been established firstly by studying pure ZnSb to test the agreement between theoretical results and experimental data and will be discussed further in the text.

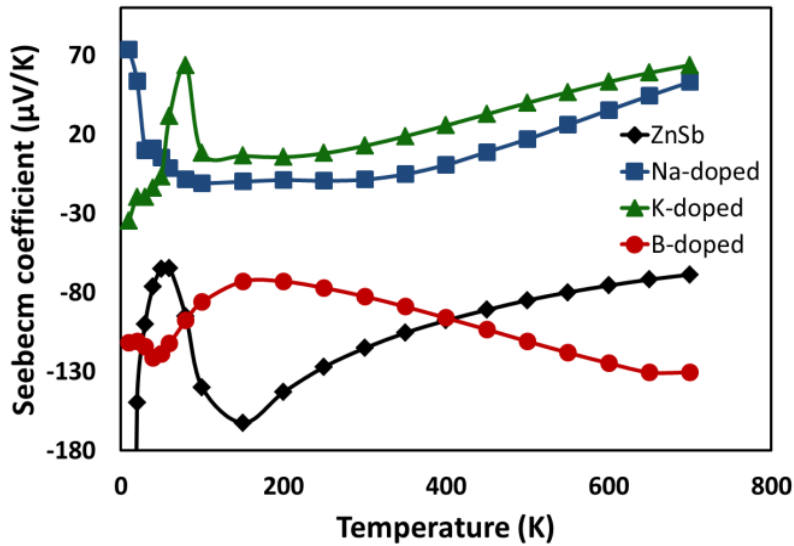

Figure 3. Calculated Seebeck coefficient as a function of temperature.

Figure 3 displays the temperature dependence between 10 and 700K of the calculated Seebeck coefficient for all considered ZnSb systems. As can be seen, two of them indicate a p-type conductivity ($NaZn_{63}Sb_{64}$ and $KZn_{63}Sb_{64}$ doped structures) and two a n-type conductivity ($BZn_{63}Sb_{64}$ and pure ZnSb). The fact that pure ZnSb shows theoretically an n-type conductivity has already been discussed[26] : the experimentally observed p-type conductivity is most probably due to the presence of intrinsic defects in the ZnSb structure, and especially of the most probable defect: Zn vacancies. The tendency of the temperature dependence of the absolute value of the Seebeck coefficient for the doped samples is similar above approximately 150K: it increases with increasing temperature, while the pure ZnSb structure shows an opposite behavior. In this case the Seebeck coefficient decreases steadily going into the direction of a sign change from positive to negative, but not reaching this point before the melting temperature.

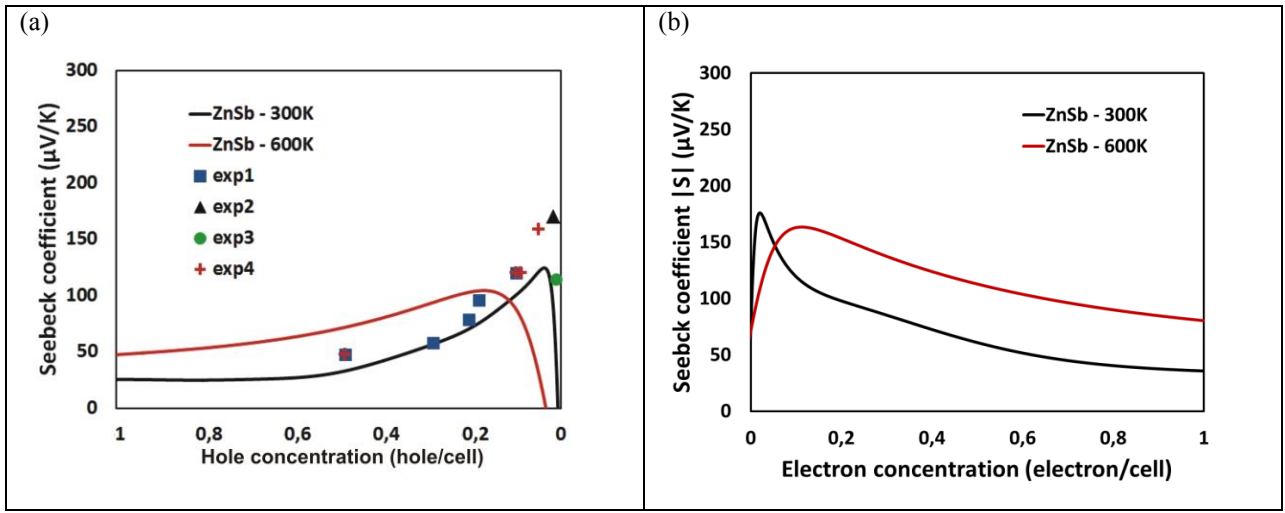

Figure 4. Seebeck coefficient as a function of the charge carrier concentration calculated at 300K and 600K for (a) p-type ZnSb along with available experimental values as found in the literature for 300K (exp1[32], exp2[33], exp3[34], exp4[35]) and (b) n-type ZnSb

The agreement between the calculated values of the Seebeck coefficient as a function of carrier concentration ($n$ indicates the electron concentration in electrons/cell (e/cell) and $p$ indicates the hole concentration in holes/cell (h/cell)) and the experimental values (figure 4(a)) indicates that the rigid band approximation (RBA) applied in the theoretical simulations is reasonable, especially if the doping level is kept low. This kind of approximation has been widely used for the theoretical study of thermoelectric materials[36]. Therefore calculated transport properties can, with a reasonable reliability, reflect the experiments as can be seen in figure 4.

This fact is giving us a foundation for further predictions of the optimal doping concentrations for the considered ZnSb compounds in this paper. As can be seen in figure 4 the difference between the dependency of $S(n)$ at 300K and 600K is small. At lower temperature the maximal values of $S$, 125 and 169 µV/K, appear at $p=0.0351$ h/cell and $n=0.013$ e/cell for p- and n-type conductivity respectively. For higher temperatures the maximal values of *S,* 104 and 163 µV/K, appear at *p*=0.15 h/cell and *n*=0.1 e/cell respectively.

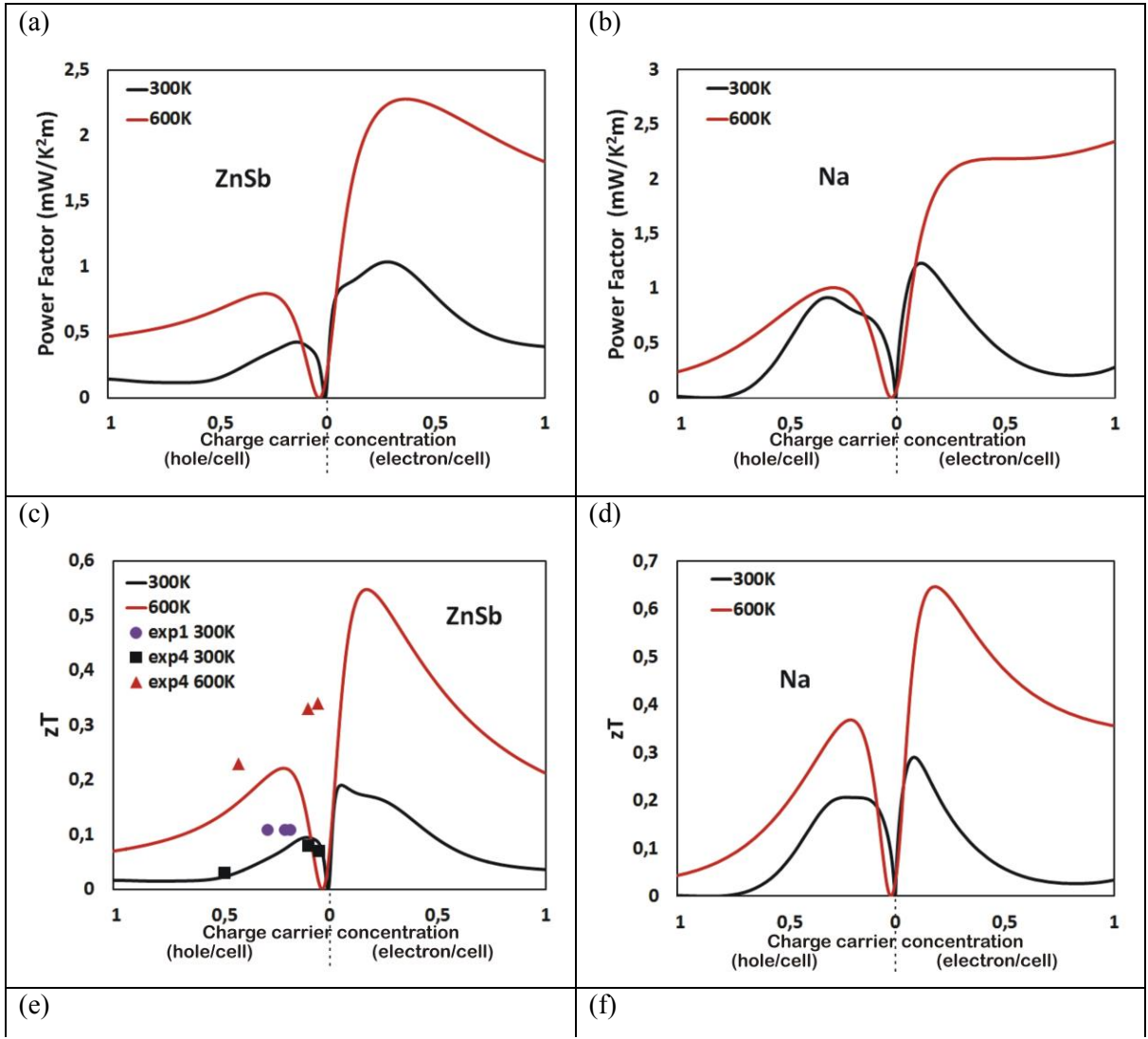

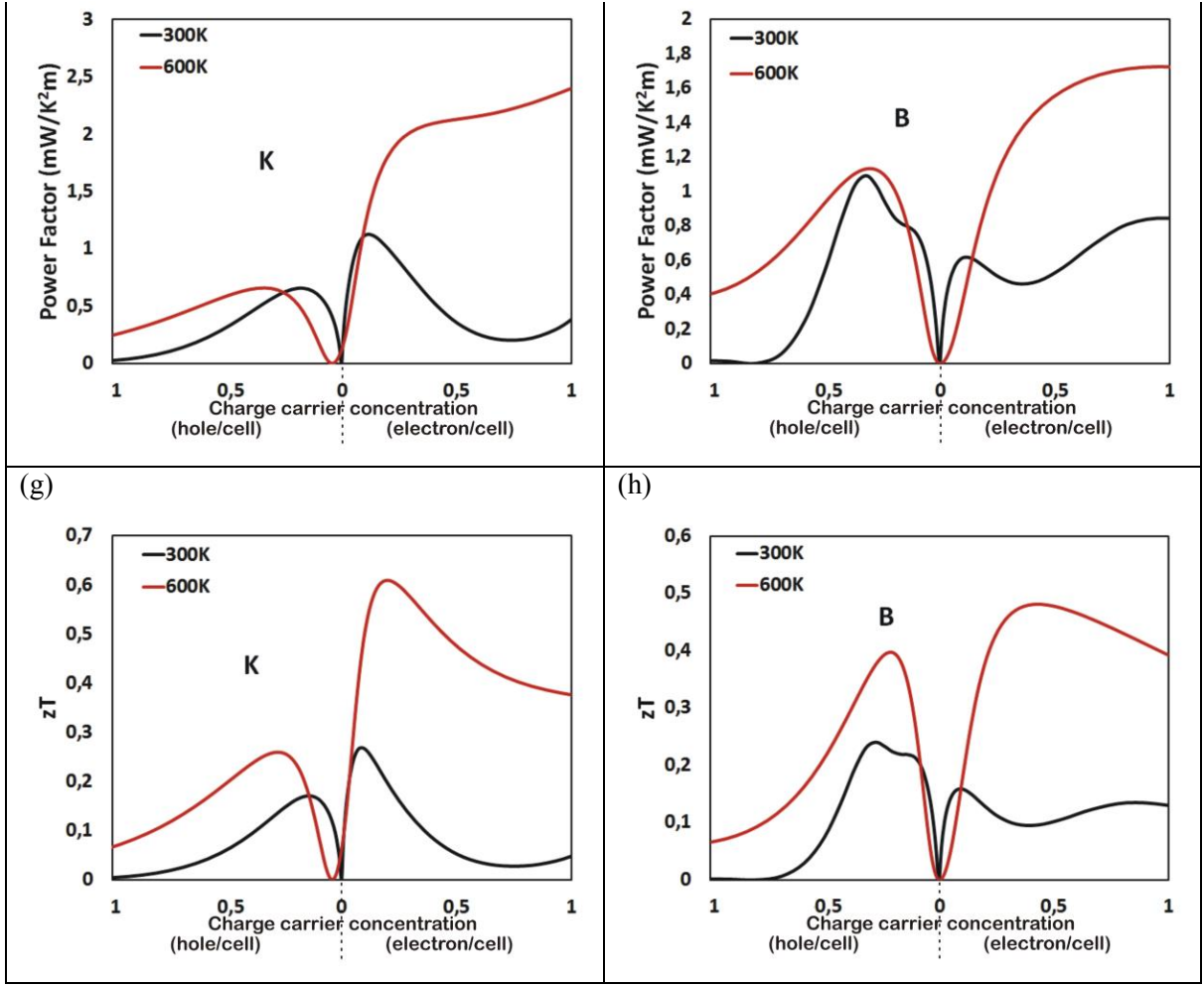

Figure 5. Calculated Power Factor ((a), (c), (e), (g)) and $zT$((b), (d), (f), (h)) as a function of charge carrier concentration (at 300K and 600K) for pure and doped ZnSb structures (AZn$_{63}$Sb$_{64}$, A=Na, K, B), along with available experimental data of $zT$ for pure ZnSb (exp1[32],exp4[35]).

In figure 5 we report the values of $S^2\sigma$ which is the Power Factor (*PF*) and the figure of merit $zT$. The figure of merit has been evaluated using the Wiedemann-Franz law to model the electronic thermal conductivity:

$$zT = \frac{S^2 \sigma/\tau}{LT \sigma/\tau + \kappa_l/\tau} T \qquad (2)$$

The Lorentz number $L$ has been taken here as a constant equal to $2.44 \cdot 10^{-8}$ W$\Omega$K$^{-2}$, the lattice part of the thermal conductivity $\kappa_l$ has been fixed to 1Wm$^{-1}$K$^{-1}$ and the constant relaxation time $\tau$ to $10^{-14}$s for reasons explained later in the text.

Figure 5(a) displays the calculated Power Factor as a function of carrier concentration. From this figure we can see that in order to maximize the p-type *PF* to 0.43 (300K) and 0.8 mWK$^{-2}$m$^{-1}$ (600K) we should obtain a ZnSb composition with $p$=0.135 (300K) and $p$=0.281(600K) h/cell. For the n-type doping the maximum PF 1.04 (300K) and 2.28 mWK$^{-2}$m$^{-1}$ (600K) can be obtained with $n$=0.288 (300K) and 0.365 e/cell (600K). The same analysis has been done for the rest of the systems and the obtained data have been gathered in table 4.

Table 4. Theoretically predicted maximal values of the Power Factor for n- and p-type ZnSb systems at 300K and 600K and the corresponding electron/cell concentrations.

|  | p-type | | | | n-type | | | |
|---|---|---|---|---|---|---|---|---|
|  | 300K | | 600K | | 300K | | 600K | |
|  | max PF | p (h/cell) | max PF | p (h/cell) | max PF | n (e/cell) | max PF | n (e/cell) |
| Pure ZnSb | 0.43 | 0.135 | 0.8 | 0.281 | 1.04 | 0.288 | 2.28 | 0.365 |
| BZn$_{63}$Sb$_{64}$ | 1.09 | 0.323 | 1.13 | 0.278 | 0.62 | 0.105 | 1.72 | 1 |
| NaZn$_{63}$Sb$_{64}$ | 0.91 | 0.299 | 1.01 | 0.304 | 1.23 | 0.126 | 2.73 | 0.174 |
| KZn$_{63}$Sb$_{64}$ | 0.66 | 0.194 | 0.66 | 0.337 | 1.12 | 0.103 | 2.73 | 1.61 |

Table 5. Theoretically predicted maximal values of the figure of merit for n- and p-type ZnSb systems at 300K and 600K and the corresponding electron/cell concentrations.

|  | p-type | | | | n-type | | | |
|---|---|---|---|---|---|---|---|---|
|  | 300K | | 600K | | 300K | | 600K | |
|  | max zT | p (h/cell) | max zT | p (h/cell) | max zT | n (e/cell) | max zT | n (e/cell) |
| Pure ZnSb | 0.095 | 0.105 | 0.22 | 0.212 | 0.19 | 0.056 | 0.54 | 0.172 |
| BZn$_{63}$Sb$_{64}$ | 0.24 | 0.279 | 0.4 | 0.201 | 0.16 | 0.105 | 0.48 | 0.405 |
| NaZn$_{63}$Sb$_{64}$ | 0.21 | 0.217 | 0.37 | 0.221 | 0.29 | 0.092 | 0.65 | 0.187 |
| KZn$_{63}$Sb$_{64}$ | 0.17 | 0.148 | 0.26 | 0.281 | 0.27 | 0.074 | 0.6 | 0.199 |

In figure 5(b) we report the calculated values of *zT* as a function of the charge carrier concentration $n/p$ for pure ZnSb. Similarly to the *PF*, *zT* shows one p-type and one n-type maximum in the concentration range from $p$=1 h/cell to $n$=1 e/cell. At 300K, a *zT* of 0.095 and 0.19 can be expected for $p$=0.105 h/cell and $n$=0.056 e/cell as well as a $zT$ of 0.22 and 0.54 for 0.212 h/cell and 0.172 e/cell at 600K. Given the approximations we used and the few experimental data that are available we believe that the agreement between theoretical predictions and experiments is satisfactory. Thus we performed the same analysis for the doped structures and the results can be found in table 5.

The experimental values of the n-type ZnSb are difficult to find since the samples are highly unstable and tend to transform into p-type. The instability of the n-type samples has been carefully investigated by A. Abou-Zeid and G. Schneider[14]. They also claim that the production of n-conducting polycrystalline ZnSb was so far not possible due to the influence of inner polycrystalline interfaces and chemisorption processes. This can be the explanation of our experimental results that will be discussed in the next part of this paper.

At this point we would like to add that the $zT$ has been calculated using several different values of the relaxation time $\tau$ (not shown), and we have chosen to display in figure 5(c) the results obtained using $\tau = 10^{-14}$ that give the best agreement between theory and experiment. This value of the relaxation time has already been widely used in $zT$ evaluations of thermoelectric materials, including in a previous study of ZnSb [37].

Considering the p-type doped structures, (NaZn$_{63}$Sb$_{64}$, KZn$_{63}$Sb$_{64}$), only a slight enhancement of the thermoelectric properties is observed (as predicted by the evaluation of the effective masses) with the concentrations considered in this paper – 0.8%at (which corresponds to 1 electron/cell), but there exists a potential to further improve the properties after tuning the concentration of the doping elements. Concerning n-type doping of ZnSb we find, through calculations, a potential new dopant: Boron (BZn$_{63}$Sb$_{64}$). This is a priori an interesting result but before moving to the laboratory we have calculated the different formation energies for the different dopants.

### 3.4. Formation energy

The enthalpies of formation $\Delta_f H$ of pure and doped structures have been calculated from the total energy data through the following equation in order to understand the macroscopic observations:

$$\Delta_f H(A_x Zn_y Sb_z) = E(A_x Zn_y Sb_z) - \left( \frac{xE(A)}{x+y+z} + \frac{yE(Zn)}{x+y+z} + \frac{zE(Sb)}{x+y+z} \right) \quad (3)$$

where the E's are the calculated total energies (in eV/atom) of the structures: $A_x Zn_y Sb_z$ and of the pure elements A, (A=Na, K, B), Zn, Sb each fully relaxed to their equilibrium geometries (see table 6). The obtained values are listed in table 7. The experimental value of the enthalpy of formation of ZnSb is relatively smaller (from -0.07 to -0.09 eV/atom)[28, 38] than the calculated one, which allows us to suspect that the overestimation is also true for the doped structures. This would mitigate the impact of our relatively elevated values. Despite the high formation energy of the Boron defect we have decided to transfer this compound to the laboratory since it would be an important step forward if such an n-type ZnSb compound could be synthesized.

Table 6. List of the equilibrium geometries used for the relaxation calculations.

| Structures | Equilibrium geometries |
|---|---|
| $A_x Zn_y Sb_z$ | Orthorhombic (Pbca) |
| Na | Bcc (Im-3m) |
| K | Bcc (Im-3m) |
| B | Rhombohedral (R-3m) |
| Zn | Hcp (P6$_3$/mmc) |
| Sb | Hexagonal (R-3m) |

Table 7. Formation enthalpies of pure and doped structures.

| Compound | Formation energy (eV/atom) | Formation energy of a defect (eV/defect) |
|---|---|---|
| $Zn_{64}Sb_{64}$ | -0.031 | - |
| $NaZn_{63}Sb_{64}$ | 0.036 | 8.67 |
| $KZn_{63}Sb_{64}$ | 0.042 | 9.33 |
| *$BZn_{63}Sb_{64}$* | *0.058* | *11.47* |

### *3.5. Macroscopic observation of the ingot*

Synthesized undoped ZnSb (0 at% of boron) looks homogeneous from the top to the bottom of the ingot. Its aspect is metallic and very well crystallized (figure 6(a)). When a melted ZnSb sample containing small amounts of boron (1 at%) was quenched, two parts in the ingot were clearly evidenced: the bottom part of the ingot presents a bulk aspect whereas in the top of the ingot black-powder is present (figure 6(b)).

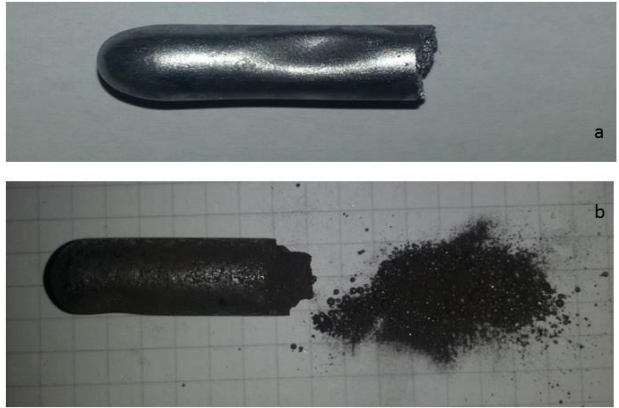

Figure 6. Macroscopic observation of (a) undoped ZnSb sample and (b) 1 at% B doped ZnSb ingot.

*3.6. Structural characterization by X-ray diffraction and Scanning Electron Microscopy (SEM)*

The X-ray diffraction patterns of ZnSb before and after annealing at 573 K are shown in figure 7. The main phase is ZnSb, moreover the peaks of Sb and $Zn_4Sb_3$ phases are only observed for quenched samples. For the sample with 1at% B ($Zn_{0.99}SbB_{0.01}$) the X-ray diffraction patterns of the milled ingot and the black powder are presented in figure 8(a) and 8(b). For the ingot (figure 8(a)), ZnSb was identified as the dominant phase. Only small peaks of $Zn_4Sb_3$ and Sb are identified in the background. figure 8(b) shows the X-ray pattern obtained for the powder. The peaks corresponding to $Zn_4Sb_3$, Sb and Zn phases are more intense. In both cases no trace of Boron appears since it was introduced in the mixture in an amorphous state.

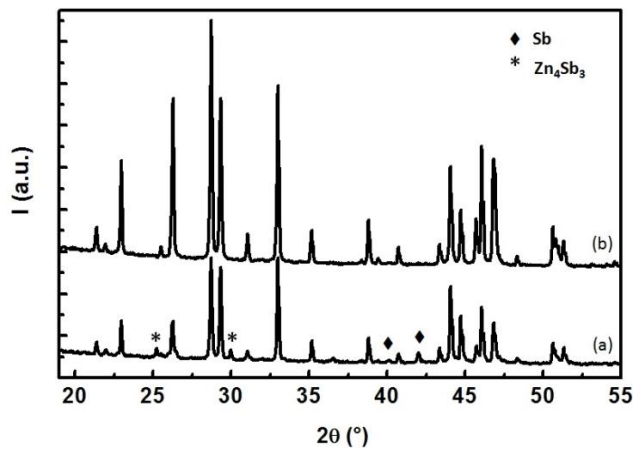

Figure 7. X-ray diffractogram of the ZnSb sample:(a) before annealing (b) after annealing (ZnSb corresponds to the unmarked peaks.

a)

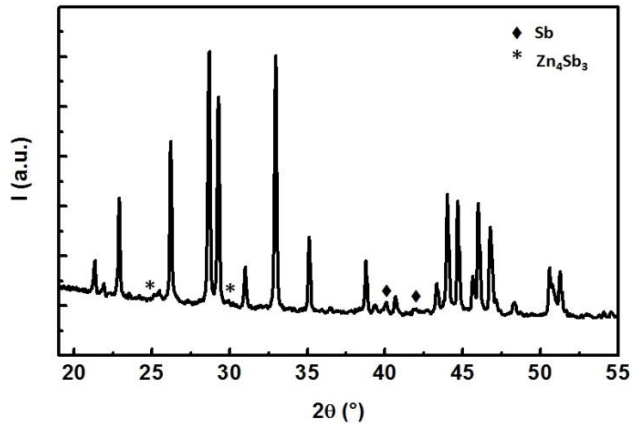

b)

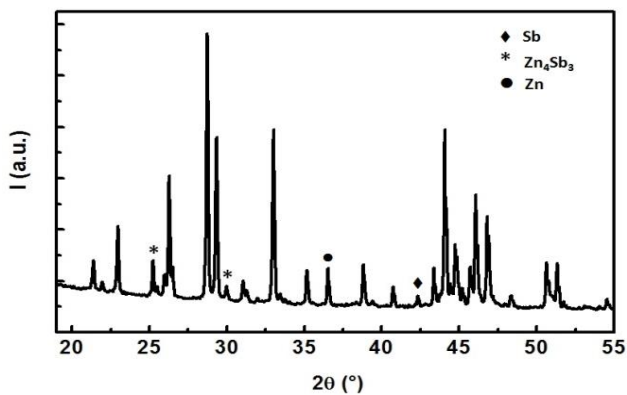

Figure 8. X-ray diffractogram of $Zn_{0.99}SbB_{0.01}$ (a) ingot and (b) black powder.

The SEM analysis of the $Zn_{0.99}SbB_{0.01}$ sample revealed two different areas. The main constituents are ZnSb grains shown in light grey. The precipitates (white area) have been identified by the combined use of EDS and SEM analysis. They have an appearance suggesting an eutectic-like reaction between the ZnSb and the Sb phases. In order to find the localization of Boron in the material we investigated the materials by EDS analysis. For that, two parts of the ingot were examined: the first part concerns the bottom of the ingot.

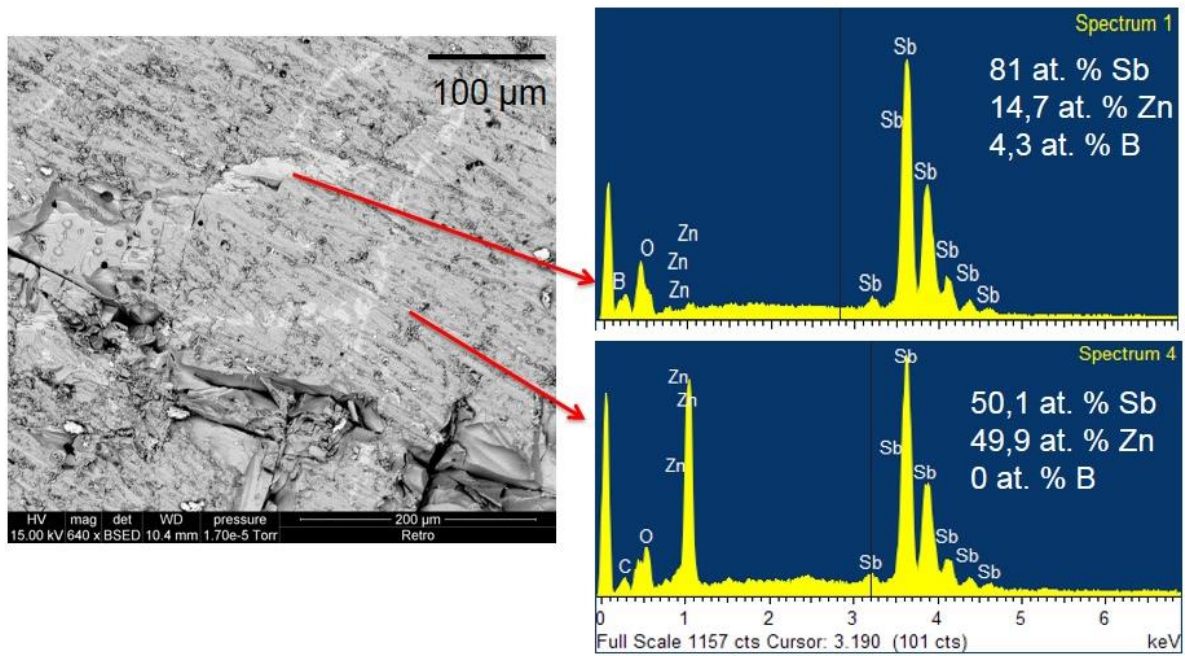

Figure 9. Scanning Electron Microscopy (SEM) observation in Back Scattering Electron Diffraction (BSED) and EDS analysis of the bottom of the ingot.

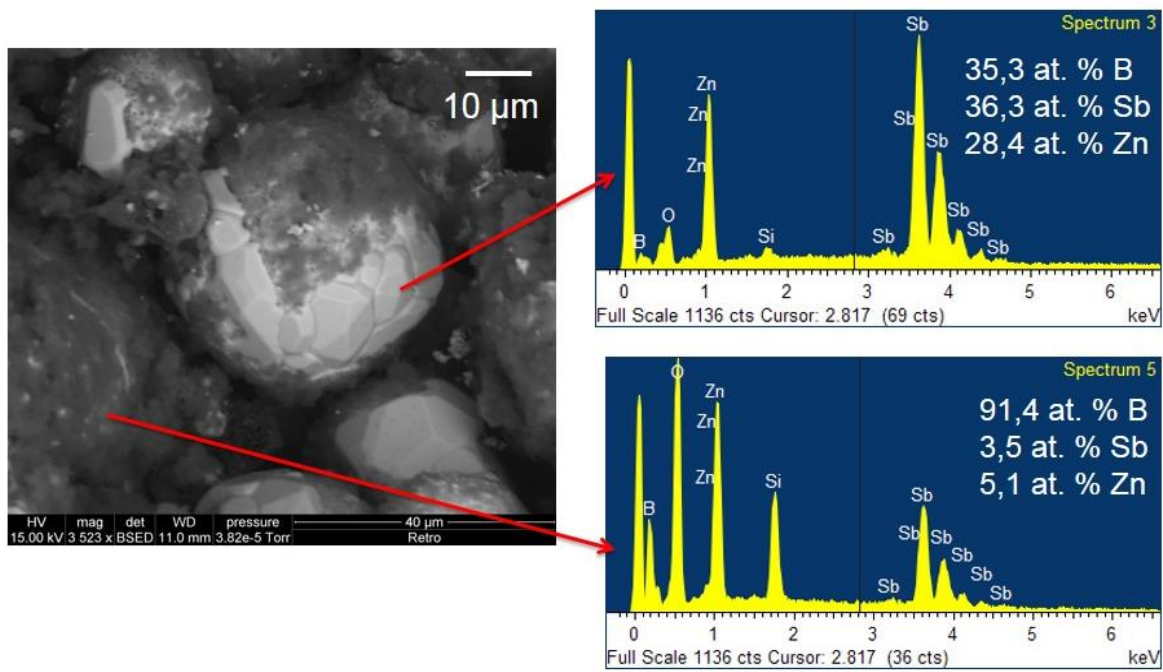

Figure 10. Scanning Electron Microscopy (SEM) observation in Back Scattering Electron Diffraction (BSED) and EDS analysis of the black powder at the top of the ingot.

Figure 9 shows the SEM micrographs obtained for these two samples associated with EDS analysis. We can distinguish a white phase located at the grain boundaries and a grey phase: spectra obtained by EDS show that Boron could be present in the white area (4.3 at%) whereas in the grey area no Boron is detected. The second analysis was performed on the black-powder located in the top of the ingot (figure 10).Two zones are clearly identified: a well crystallized phase with a metallic aspect which indubitably corresponds to the ZnSb phase and the black powder with a high percentage of Boron in it (70.6 at% of Boron).

These experimental results show that Boron is not introduced into the structure of ZnSb but seems to be present at the grain boundaries with Antimony. Our phase identification studies indicate that the solubility of B in ZnSb is low. Moreover, the ingot obtained by solidification is not homogeneous. Furthermore, the concentration of Boron in the sample increases from the bottom to the top of the ingot. A possible explanation of this phenomenon is that at the beginning of the process, Sb, Zn and B were mixed together and the repartition of the three elements was homogeneous in the ampoule. Afterwards when this mixture was heated and melted in the furnace, Boron could not be incorporated in the structure of ZnSb, possibly due to the high formation enthalpy of a substitutional B defect (11.47eV/defect – see table 7). Then, as the density of boron ( $d_B^{(sol)}$= 2.8 g/cm³) is very low compared to that of Sb ( $d_{Sb}^{(liq)}$= 6.53 g/cm³) and Zn ($d_{Zn}^{(liq)}$ = 6.57 g/cm³), the Boron migrates to the top of the sample in the liquid state. After quenching, the ingot appears in two parts: the bottom composed of ZnSb + (Sb,B eutectic) in small proportions and the top with a black powder essentially composed of Boron. In order to try to homogenize this sample, several thermal treatments were realized. In each case, the results give rise to a mixture of ZnSb compound + (Sb, B eutectic).

4. Conclusions

In the present paper we have shown a theoretical and experimental work on orthorhombic ZnSb doped with Na, K and B with the aim of designing better thermoelectric compounds. We have studied the electronic structure of these systems and shown the similarity of their band structures and determined their effective masses which compare well with previously published experimental data. Based on the analysis of the deformation charge densities we have obtained the character of the bonds between the elements of the ZnSb matrix and the doping elements. Our first principles investigation has shown that the substitution of zinc by sodium and potassium results in p-type compounds with a maximum figure of merit of 0.21 and 0.17 (0.37 and 0.26) respectively at 300K and 600K. Presence of Boron in the ZnSb matrix leads to n-type conductivity – which is a new result – with a zT of 0.16 and 0.48 at 300K and 600K respectively.

Unfortunately for the moment the synthesis of this compound seems to be unreachable due to its high formation energy. Nevertheless the boron doped structure has a high potential in the production of an n-type leg in a ZnSb based thermoelectric device if the formation energy barrier can be overcome.

**Acknowledgements:** We thank the companies Total and Hutchinson for supporting financially our work on the development of ZnSb based thermoelectric materials. J.C Tedenac is acknowledged for fruitful discussions. Part of the calculations have been performed at the "Centre Informatique National de l'Enseignement Supérieur" (CINES) computer centre in Montpellier.